\documentclass[a4paper,11pt,times,authoryears]{article}
\usepackage{authblk}
\usepackage[square,numbers]{natbib}
\usepackage{tikz}
\usepackage{listings}
\usepackage{bm}
\usepackage{amsmath}
\usepackage{amsthm}
\usepackage{mathtools}
\usepackage{amssymb,amsfonts}
\usepackage{graphicx}
\usepackage{mathrsfs}
\usepackage{multirow, multicol}
\usepackage{enumitem, array}
\usepackage{subcaption}
\usepackage{epstopdf}
\usepackage{float}
\usepackage{enumerate}
\usepackage{booktabs}
\usepackage{setspace}
\usepackage{adjustbox}
\usepackage{pdflscape}
\usepackage{verbatim}

\usepackage[colorlinks,citecolor=blue]{hyperref}
\usepackage{rotating}
\usepackage{fullpage}
\usepackage{setspace}
\usepackage{lscape}
\usepackage{booktabs}
\usepackage{setspace}
\usepackage{tocbibind}
\usepackage{color}
\usepackage{commath} 
\usepackage{algorithm}
\usepackage{rotating}
\usepackage[noend]{algpseudocode}

\newtheorem{Theorem}{Theorem}[section]

\newtheorem{proposition}[Theorem]{Proposition}
\usepackage{geometry}
\begin{document}

\title{Exploring Discrete Factor Analysis with the discFA Package in R}
\author{{Reza Arabi Belaghi$^\dag$, Yasin Asar$^{\ddag}$ and Rolf Larsson$^{*}$ } \vspace{.5cm} 
\\$^{\dag}$Swedish Agriculture University (SLU), Uppsala, Sweden\\
e-mail: rezaarabi11@gmail.com
\\
$^{\ddag}$Corresponding Author\\Department of Mathematics and Computer Sciences,\\ Necmettin Erbakan University, Konya, Turkey\\
e-mail: yasar@erbakan.edu.tr, yasinasar@hotmail.com\\
$^{*}$Uppsala University, Uppsala, Sweden\\rolf.larsson@math.uu.se}

\date{}
\maketitle
\begin{abstract}
Literature suggested that using the traditional factor analysis for the count data may be inappropriate. With that in mind, discrete factor analysis builds on fitting systems of dependent discrete random variables to data. The data should be in the form of non-negative counts. Data may also be truncated at some positive integer value. The \href{https://cran.r-project.org/web/packages/discFA/index.html}{\texttt{discFA}} package in \texttt{R} allows for two distributions: Poisson and Negative Binomial, in combination with possible zero inflation and possible truncation, hence, eight different alternatives. A forward search algorithm is employed to find the model optimal factor model with the lowest AIC. Several  different illustrative examples from psychology, agriculture, car industry, and a simulated data will be analyzed at the end. 
\bigskip
\\
\noindent{\bf Keywords}: Discrete Factor Analysis, AIC, Poisson, Negative Binomial, Zero-inflated, Model Selection, discFA, R package

\end{abstract}


\onehalfspacing
\noindent
\section[Introduction]{Introduction} \label{sec:intro}
Traditional factor analysis, while a powerful tool, often proves inappropriate for analyzing count data. This is primarily because count data, characterized by non-negative integer values, frequently violate the assumptions of normality and continuous measurement inherent in classical factor analytic approaches. Recognizing this limitation, discrete factor analysis (DFA) offers a robust alternative by modeling systems of dependent discrete random variables directly.

It is well known that the traditional factor analysis method was build on normality assumptions \citep{joreskog2016multivariate} and, specifically, the analysis of the empirical covariance matrix of the data. In particular, ordinal data can be considered, by assuming underlying normal random variables and estimating thresholds. Along these lines, a full likelihood approach is numerically very complicated  as well as time consuming. 
Literature suggests that it may not be appropriate  to treat nominal and ordinal data as
interval or ratio \citep{basto2012spss}. Applying
traditional factor analysis procedures to item-level data almost always produces misleading
results. Several authors explain the scale problem and suggest alternative procedures when
attempting to establish the validity of a Likert scale since ordinal variables do not have a
metric scale and tend to be attenuated due to the restriction on range \citep{marcus1987meaningless, mislevy1986recent}. Also, traditional factor analysis procedures produce meaningful results only if the data is continuous and is also
multivariate normal. Discrete data almost never meet these requirements \citep{basto2012spss}. 

A simpler yet very efficient way to handle discrete factor analysis for count data, and in particular ordinal data, is proposed by \citet{larsson2020discrete}. Here, the approach is to form a system of dependent variables by explicit use of additive common factors. For count data, \citet{larsson2020discrete} uses the Poisson distribution, and truncates to analyze ordinal data. The method is further applied by \citet{larsson2021applications}, who also extend it to deal with over dispersion via the negative binomial distribution.

To enhance the applicability of the method, the present paper presents a newly developed \texttt{R} package which is called {\texttt{discFA}}. The \texttt{discFA} package in \texttt{R} provides a flexible framework for implementing DFA, offering eight distinct model alternatives. These alternatives stem from the combination of two primary distributions—Poisson and negative binomial—with the additional considerations of zero-inflation and truncation. The Poisson distribution is suitable for count data where the variance is approximately equal to the mean, while the negative binomial distribution is more appropriate when over-dispersion (variance greater than the mean) is present, which is often the case with real-world count data. The inclusion of zero-inflation addresses situations where there is an excess of zero counts beyond what the chosen distribution would predict.

To identify the most parsimonious and well-fitting model, the \texttt{discFA} package employs a forward search algorithm. This algorithm systematically explores different factor structures, ultimately selecting the optimal model based on the lowest Akaike Information Criterion (AIC). AIC is a widely used measure for model selection, balancing model fit with model complexity.

The utility and versatility of DFA will be demonstrated through several illustrative examples. These examples, drawn from diverse fields such as psychology, agriculture, and the automotive industry, along with an analysis of simulated data, will highlight the practical application and advantages of DFA in handling complex count data structures.

The rest of the paper is as follows. Section \ref{sec:models} presents the new methodology for the discrete factor analysis. In Section \ref{sec:discFA}, the new \texttt{R} package \texttt{discFA} is described. Real data analysis and applications are given in Section \ref{sec:illustrations}. The limitations of the package are discussed in section   \ref{sec:limit}. Section \ref{sec:summary} concludes the paper.

\section{New methodology} \label{sec:models}

\subsection{Theory of discrete factor analysis}
One way to construct a pair of dependent random variables, that may be discrete, is to form the system
\begin{equation}
\left\{\begin{array}{rcl}
Y_1&=&U+X_1,\\
Y_2&=&U+X_2,\end{array}\right.\label{Karlis}
\end{equation}
where $U$, $X_1$ and $X_2$ are simultaneously independent random variables. The variables $U$, $X_1$ and $X_2$ are considered latent, only $Y_1$ and $Y_2$ are observed. An interesting special case  is when $U$, $X_1$ and $X_2$ are Poisson distributed, implying that $Y_1$ and $Y_2$ are also Poisson distributed (\citet{karlis2003algorithm}). As it is easily seen, the covariance between $Y_1$ and $Y_2$ equals the variance of $U$. Hence, negative correlation between $Y_1$ and $Y_2$ is not allowed.

It is straightforward to construct the likelihood function from \eqref{Karlis}. Say that the distributions of $U$, $X_1$ and $X_2$ are discrete and have parameters $\theta_0$, $\theta_1$ and $\theta_2$, respectively, possibly vector valued. Denote the corresponding probability mass functions (pmf) by $f(u;\theta_j)$, for $j=0,1,2$.
Suppose we have observation pairs $(y_{11},y_{12}),\ldots,(y_{n1},y_{n2})$. Then, the likelihood takes the form
\begin{equation}
L(\theta_0,\theta_1,\theta_2)=\prod_{i=1}^n\sum_{u_i=0}^{\min(y_{i1},y_{i2})} 
f(u_i;\theta_0)f(y_{i1}-u_i;\theta_1)f(y_{i2}-u_i;\theta_2),\label{Bivlik}
\end{equation}
where we used that $Y_1$ and $Y_2$ are conditionally independent, given $U=u$. 

It is straightforward to generalize \eqref{Karlis} to any number of $Y_j$ variables, say $m$, by writing $Y_j=U+X_j$ for $j=1,...,m$, where $U,X_1,...,X_m$ are simultaneously independent. The likelihood equation \eqref{Bivlik} generalizes accordingly as
\begin{align}
L(\theta_0,\theta_1,...,\theta_m)
=\prod_{i=1}^n\sum_{u_i=0}^{\min(y_{i1},...,y_{im})} 
f(u_i;\theta_0)f(y_{i1}-u_i;\theta_1)\cdot...\cdot f(y_{im}-u_i;\theta_m),\label{mlik}
\end{align}
where we have $m$-tuples of observations $(y_{11},...,y_{1m}),\ldots,(y_{n1},,...,y_{nm})$, and where the distributions of $U,X_1,...,X_m$ have parameters (possibly vectors) $\theta_0,\theta_1,...,\theta_m$.

After specification of the pmf $f$, the likelihood in \eqref{mlik}, or the log of it, may be readily maximized by using numerical methods.

The package uses two basic distributions, the Poisson distribution with the following pmf
\begin{equation}
f(x;\theta)=\frac{\theta^x}{x!}e^{-\theta},\label{po}
\end{equation}
where $\theta>0$, and the negative binomial distribution with the following pmf
\begin{equation}
f(x;\theta)=\binom{r+x-1}{x}p^r(1-p)^x,\label{negbin}
\end{equation}
where $\theta=(r,p)$ with $r>0$ and $0<p<1$, and in both cases, $x=0,1,...$. Both are natural models for count data, where in the Poisson case, the expectation equals the variance, while negative binomial is over dispersed, i.e. with the variance is greater than the expectation.

A modification of these distributions is zero inflatedness, where we put an extra probability mass at zero so that for some $0<\pi<1$, the pmf becomes
\begin{equation}
f^{(Z)}(x,\theta)=\pi I\{x=0\}+(1-\pi)f(x;\theta),\label{zi}
\end{equation}
where $I\{x=0\}$ is one if $x=0$ and zero otherwise, and $f(x;\theta)$ is the pmf for the original distribution.

In case the data is truncated, say at $A$, (\ref{mlik}) becomes
\begin{equation}
L(\theta_0,\theta_1,\ldots,\theta_m)
=\prod_{i=1}^n\frac{\sum_{u_i=0}^{\min(y_{i1},\ldots,y_{im})} 
f(u_i;\theta_0)f(y_{i1}-u_i;\theta_1) \ldots f(y_{im}-u_i;\theta_m)}
{\sum_{u=0}^A\sum_{y_1=0}^{A-u}\ldots\sum_{y_m=0}^{A-u} 
f(u;\theta_0)f(y_{1}-u;\theta_1) \ldots f(y_{m}-u;\theta_m)}
.\label{trunc}
\end{equation}

If the distribution belongs to the exponential family, which for our package corresponds to the plain Poisson and negative binomial cases, numerical maximization of the (log) likelihood is simplified because the estimated mean of the distribution can be shown always to equal the empirical mean of the observations. In practice, this means that in a system of $m$ dependent Poisson variables, numerical maximization only needs to be performed over the $U$ parameter $\theta_0$. In the negative binomial case with $m$ variables, the maximization may e.g. be performed only with respect to the parameters $r_0,p_0,p_1, \ldots, p_m$, where $\theta_j=(r_j,p_j)$ for $j=0,1, \ldots, m$.

Indeed, we have the following proposition:

\begin{proposition}
Let $Y_j=U+X_j$ for $j=1,\ldots,m$, where the random variables $U,X_1,\ldots,X_m$ are simultaneously independent, with probability mass functions of the form
$$f(x;\theta)=h(x)\exp\{xa(\theta)-b(\theta)\},$$
where $a(\theta)$ is not constant in $\theta$.
Then, when the parameters equal their respective MLEs, for all $j$, $E(Y_j)=\bar y_j$, where $\bar y_j$ is the sample mean for variable $j$.
\end{proposition}
See the \textbf{appendix} for proof.

\subsection{Model selection}
In the present paper, following \citet{larsson2020discrete}, the proposed procedure tries to find the model with the smallest value of the Akaike Iformation Criterion (AIC), see  \citet{akaike1974new}. We will use the following definition
\begin{equation}
{\rm AIC}=-2\log L_{max}+2p,\label{AIC}
\end{equation}
where $L_{max}$ is the maximum likelihood value and $p$ is the number of parameters.
 
However, given the number of variables $N$, there are extremely many ways to form a factor model. In \citet{larsson2020discrete}, since the number of possible models grows exponentially with $N$, a forward selection procedure is proposed. In the Poisson and truncated Poisson cases for $N=5$ and 7, this procedure is evaluated in a simulation study. It was found that in most cases, the procedure selects the model with the lowest AIC with high probability.

The selection procedure is step-wise as follows. The principle in all steps is to take the favorite model of the previous step and then merge any two groups (considering the ones to be groups of their own). The first step is to take the independence model (1,1,...,1), and compare it with all possible (2,1,...,1) models. The algorithm stops if the independence model has the lowest AIC. If not, in the next step we check all (3,1,...,1) models where the three variable group contains the pair of variables that were in the same group in the first step, together with all (2,2,1,...,1) models where we add a new pair of variables that consists of any two that were not in the first pair. If none of these models is better than the previously chosen (2,1,...,1) model, we stop and choose the previous model. If not, we go on to test new models, and so it goes on. 

The following figure  depicts the model selection procedure in dimension five.
\begin{small}
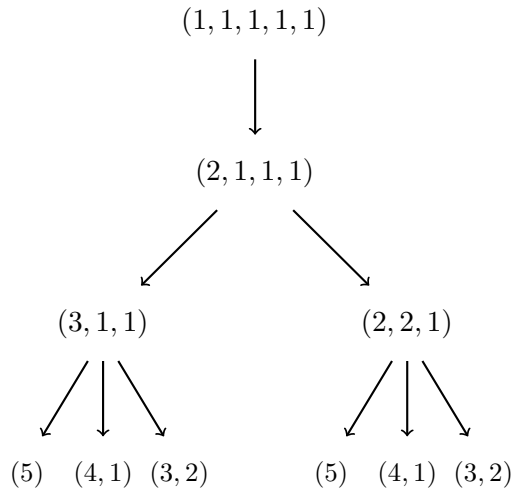
\begin{figure}[ht] \label{fig11}
\begin{center}
\begin{tikzpicture}
\draw (0,6) node {$(1,1,1,1,1)$};
\draw (0,4) node {$(2,1,1,1)$};
\draw [->,thick] (0,5.5) -- (0,4.5);
\draw (-2,2) node {$(3,1,1)$};
\draw (2,2) node {$(2,2,1)$};
\draw [->,thick] (-0.5,3.5) -- (-1.5,2.5);
\draw [->,thick] (0.5,3.5) -- (1.5,2.5);
\draw (-3,0) node {\small$(5)$};
\draw (-2,0) node {\small$(4,1)$};
\draw (-1,0) node {\small$(3,2)$};
\draw (1,0) node {\small$(5)$};
\draw (2,0) node {\small$(4,1)$};
\draw (3,0) node {\small$(3,2)$};
\draw [->,thick] (-2.2,1.5) -- (-2.8,0.5);
\draw [->,thick] (-2,1.5) -- (-2,0.5);
\draw [->,thick] (-1.8,1.5) -- (-1.2,0.5);
\draw [->,thick] (1.8,1.5) -- (1.2,0.5);
\draw [->,thick] (2,1.5) -- (2,0.5);
\draw [->,thick] (2.2,1.5) -- (2.8,0.5);
\end{tikzpicture}
\caption{Model selection algorithm for dimension 5.}

\end{center}
\end{figure}
\end{small}

Note that all optimization and calculations have been done simply by the \texttt{nlminb()} function in the \texttt{stats} package in \texttt{R}. 

\section{discFA as a New R Package}\label{sec:discFA}

In this section, we describe \texttt{discFA} package with eight different models for various count data models. We start with the Poisson and negative binomial distributions. We then add the truncated and zero-inflated models to create more flexibly for various data analysis.  Table \ref{table1} summarizes all the eight functions developed for this package. As it is clear from Table \ref{table1}, the syntax for the \texttt{discFA} is quite easy to use. Also for each model, the AIC of the best model (lowest AIC), estimated models and factor parameters are displayed. 
\begin{table}[ht] 
\tiny
\caption{Models and outputs in discFA package}
    \label{table1}
    \centering
    \begin{tabular}{l|l|l|l}
       Model  & function name& argument(s) & output  \\ \hline
        Poisson & dfp()&data&AIC,Best Model, estimated parameters\\ \hline
        Truncated Poisson & dfpt()&data and truncation &AIC,Best Model, estimated parameters\\ \hline
        Zero-inflated Poisson & dfzip()&data  &AIC,Best Model, estimated parameters\\ \hline
        Zero-inflated Truncated Poisson & dfzipt()&data and truncation &AIC,Best Model, estimated parameters\\ \hline
        Negative Binomial & dfnb()&data&AIC,Best Model, estimated parameters\\ \hline
        Truncated Negative Binomial & dfnbt()&data and truncation &AIC,Best Model, estimated parameters\\ \hline
        Zero-inflated Negative Binomial & dfzinb()&data  &AIC,Best Model, estimated parameters\\ \hline
        Zero-inflated Truncated Negative Binomial & dfzinbt()&data and truncation &AIC,Best Model, estimated parameters\\ \hline
            \end{tabular}
\end{table}

To install the package in \texttt{R}, one can simply use the following code as the usual way for other packages:

\noindent\textbf{Example 1.}
\begin{lstlisting}[language=R]
R> install.packages("discFA")
\end{lstlisting}
and then call it with the following function\\ 
\noindent\textbf{Example 2.}
\begin{lstlisting}[language=R]
R> library(discFA)
\end{lstlisting}

\section{Data Analysis and Examples} \label{sec:illustrations}
In this part, we have analyzed several  real examples from various disciplines to show to show how the \texttt{discFA} works. Some of the datasets are included in the package and the rest are available online.   

\subsection{Example 1: Simulated Data set}
\begin{figure}[h]
  \centering
  \includegraphics[scale=.5]{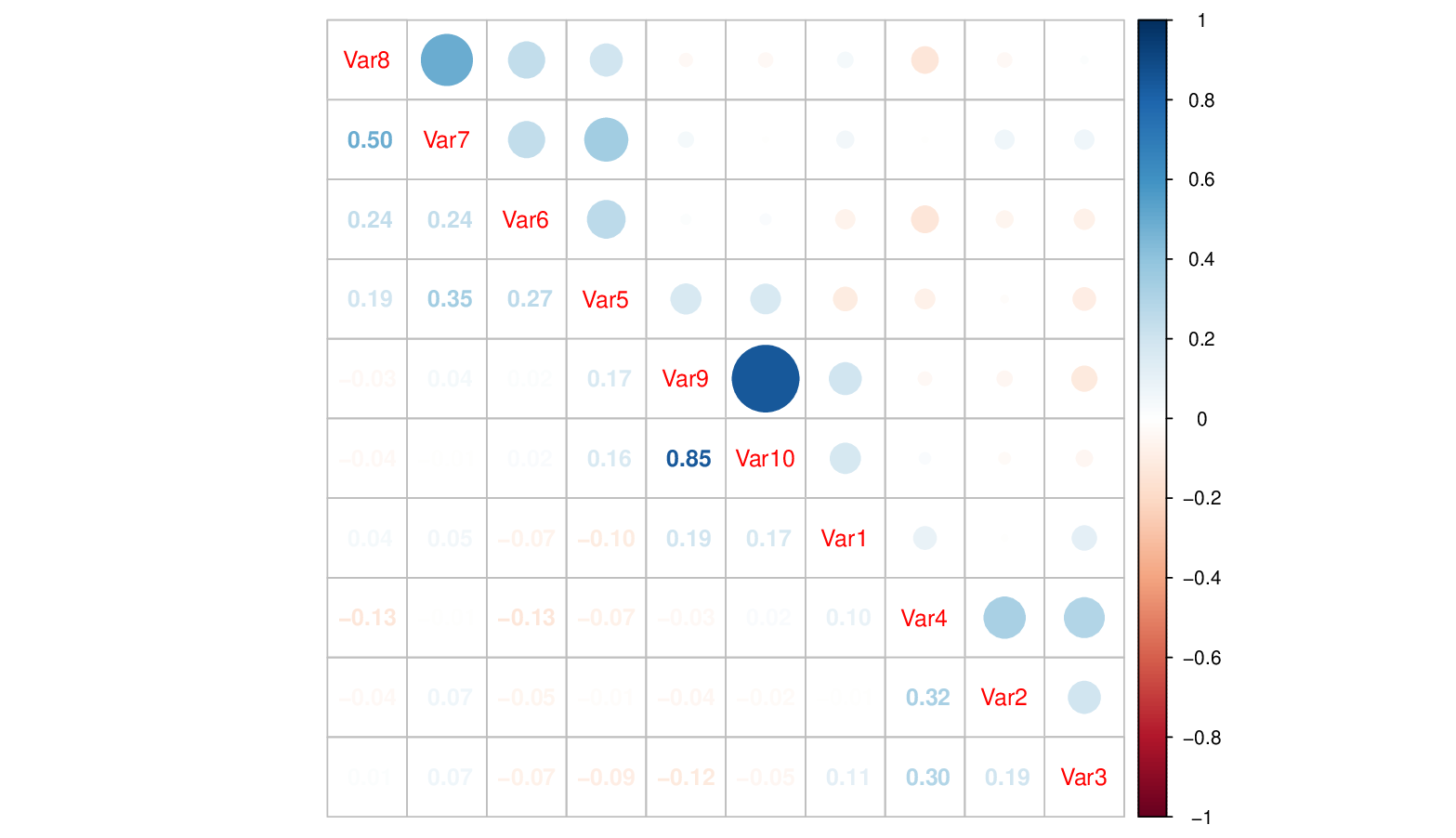}
\caption{Correlation plot for the simulated data set}
    \label{fig:zeroinlated data_dist}
\end{figure}

Here, we simulated a dataset with 10 variables form a truncated zero-inflated negative binomial  ($30\%$ zero-inflation rate) on the set of $\{0,1,2,3,4,5,6\}$. The true  model is $(1,3,4,2)$. The dataset is available in the \texttt{discFA} with the name \texttt{zinb100\_Data}. The correlation plot for this data is given in Figure \ref {fig:zeroinlated data_dist}. For the sake of comparison, we apply all models (available in the package) including the zero-inflated Poisson and negative binomial models. The results are provided as the following.

First, we start with the \textbf{Poisson model} and apply it by the following code:
\begin{lstlisting}[language=R]
R> dfp(zinb100_Data)
\end{lstlisting}
The results for this model are given below:
\begin{lstlisting}[language=R]
Call:
dfp(y = zinb100_Data)

This is a (1, 3, 4, 2) model.

AIC value is 31.50432.

Factors and variables in each factor:
  Factor1 Factor2 Factor3 Factor4
1 Var1    Var2    Var5    Var9   
2         Var4    Var7    Var10  
3         Var3    Var8           
4                 Var6           

Estimated parameters for each variable within each factor:
  Factor1 Factor2 Factor3 Factor4
1 0.84    1.6014  0.9574  0.1443 
2         0.7614  0.2874  0.5943 
3         1.3914  0.6774         
4                 1.4774         

Estimated parameters for factors:
0.4286 0.3526 1.7557

Timing:
Time difference of 2.523 secs
\end{lstlisting}

Next, we proceed with \textbf{the truncated Poisson model} using the following code:

\begin{lstlisting}[language=R]
R> dfpt(zinb100_Data,6) 
\end{lstlisting}
The results are given below:
\begin{lstlisting}[language=R]
Call:
dfpt(y = zinb100_Data, A = 6)

This is a (1, 1, 1, 1, 1, 1, 2, 2) model.

AIC value is 32.8958.

Factors and variables in each factor:
  Factor1 Factor2 Factor3 Factor4 Factor5 Factor6
1 Var1    Var2    Var3    Var4    Var5    Var6   
2                                                
  Factor7 Factor8
1 Var7    Var9   
2 Var8    Var10  

Estimated parameters for each variable within each factor:
  Factor1 Factor2 Factor3 Factor4 Factor5 Factor6
1 0.8402  2.0579  1.8356  1.1914  1.3125  1.8461 
2                                                
  Factor7 Factor8
1 0.6828  1.2318 
2 1.03    2.35   

Estimated parameters for factors:
0.32 1

Timing:
Time difference of 5.909 secs
\end{lstlisting}
We observe that the truncated Poisson model provided higher AIC compared to the usual Poisson model, even though the latter is a more natural choice given the situation. One explanation might be that in the truncated case, the selection procedure fails to find the model with the lowest AIC. 

Next, we apply the 
\textbf{zero-inflated Poisson model} by the following code: 
\begin{lstlisting}[language=R]
R> dfzip(zinb100_Data)   
\end{lstlisting}
The results for this model are given below: 
\begin{lstlisting}[language=R]
Call:
dfzip(y = zinb100_Data)

This is a (1, 3, 4, 2) model.

AIC value is 29.7.

Factors and variables in each factor:
  Factor1 Factor2 Factor3 Factor4
1 Var1    Var2    Var5    Var9   
2         Var4    Var7    Var10  
3         Var3    Var8           
4                 Var6           

Estimated zero-inflated parameters for each variable within each factor:
  Factor1 Factor2 Factor3 Factor4
1 0.354   0.3527  0.5093  0.5218 
2         0.6108  0.8937  0.4067 
3         0.3097  0.5283         
4                 0.2897         

Estimated parameters for each variable within each factor:
  Factor1 Factor2 Factor3 Factor4
1 1.3003  2.1717  1.8279  0.2566 
2         1.4536  2.1342  0.9652 
3         1.7323  1.3076         
4                 1.9947         

Estimated zero-inflated parameters for each factor:
0.35 0.642 0.21

Estimated parameters for factors:
0.961 1.15 2.25

Timing:
Time difference of 1.946 mins
\end{lstlisting}
It can be easily concluded that the zero-inflated Poisson model provided a lower AIC compared to the previous models. As a result, we choose the zero-inflated Poisson model. However, we continue the analysis with the \textbf{truncated zero-inflated model} as follows: 

\begin{lstlisting}[language=R]
R> dfzipt(zinb100_Data,6)    
\end{lstlisting}
The results are provided below:
\begin{lstlisting}[language=R]
Call:
dfzipt(y = zinb100_Data, A = 6)

This is a (1, 3, 4, 2) model.

AIC value is 29.6.

Factors and variables in each factor:
  Factor1 Factor2 Factor3 Factor4
1 Var1    Var2    Var5    Var9   
2         Var4    Var7    Var10  
3         Var3    Var8           
4                 Var6           

Estimated zero-inflated parameters for each variable within each factor:
  Factor1 Factor2 Factor3 Factor4
1 0.3548  0.3558  0.5106  0.5192 
2         0.611   0.8936  0.4077 
3         0.3114  0.5288         
4                 0.292          

Estimated parameters for each variable within each factor:
  Factor1 Factor2 Factor3 Factor4
1 1.3046  2.2352  1.8516  0.2539 
2         1.4678  2.1847  0.9663 
3         1.7546  1.3121         
4                 2.0304         

Estimated zero-inflated parameters for each factor:
0.347 0.642 0.214

Estimated parameters for factors:
0.951 1.16 2.32

Timing:
Time difference of 2.984 mins
\end{lstlisting}

By comparing all four types of the Poisson models in terms of the AIC, we can easily conclude that the \textbf{zero-inflated truncated Poisson model} is the most preferable one. However, we have not examined the negative binomial models yet. Let us consider all four types of negative binomial factor analysis models that we developed in the package. We continue with the simplest model to the most complicated one: 

We start with the \textbf{negative binomial model} as follows:
\begin{lstlisting}[language=R]
R> dfnb(zinb100_Data)   
\end{lstlisting}
The results for this model are obtained as follows:

\begin{lstlisting}[language=R]
Call:
dfnb(y = zinb100_Data)

This is a (1, 3, 3, 1, 2) model.

AIC value is 30.5.

Factors and variables in each factor:
  Factor1 Factor2 Factor3 Factor4 Factor5
1 Var1    Var2    Var5    Var7    Var9   
2         Var4    Var6            Var10  
3         Var3    Var8                   

Estimated value of r for the negative binomial distributed observations(s):
  Factor1 Factor2 Factor3 Factor4 Factor5
1 1.3505  1.3441  0.6502  0.4509  0.9265 
2         0.4746  1.5289          1.1657 
3         1.6087  0.5052                 

Estimated value of p for the negative binomial distributed observations(s):
  Factor1 Factor2 Factor3 Factor4 Factor5
1 0.6165  0.4869  0.4086  0.4133  0.8842 
2         0.4515  0.5113          0.6711 
3         0.5714  0.4331                 

Estimated value of r for the negative binomial distributed factor(s):
0.994 0.275 2.64

Estimated value of r for the negative binomial distributed factor(s):
0.618 0.427 0.598

Timing:
Time difference of 1.038 mins
\end{lstlisting}

The negative binomial model provided an AIC of $30.45482$ which is lower than the Poisson model ($31.50432$). 

We proceed the example with \textbf{truncated negative binomial} model by running the following code: 

\begin{lstlisting}[language=R]
R> dfnbt(zinb100_Data,6) 
\end{lstlisting}
The results are given as follows:

\begin{lstlisting}[language=R]
This is a (1, 3, 4, 2) model.

AIC value is 29.3.

Factors and variables in each factor:
  Factor1 Factor2 Factor3 Factor4
1 Var1    Var2    Var5    Var9   
2         Var4    Var7    Var10  
3         Var3    Var8           
4                 Var6           

Estimated value of r for the negative binomial distributed observations(s):
  Factor1 Factor2 Factor3 Factor4
1 0.5337  0.7893  0.4495  1      
2         0.3751  0.0451  1      
3         1       0.4898         
4                 0.9602         

Estimated value of p for the negative binomial distributed observations(s):
  Factor1 Factor2 Factor3 Factor4
1 1       0.2978  0.2835  0.8927 
2         0.372   0.0394  0.6344 
3         0.4258  0.425          
4                 0.3571         

Estimated value of r for the negative binomial distributed factor(s):
0.971 0.323 1

Estimated value of p for the negative binomial distributed factor(s):
0.607 0.425 0.283

Timing:
Time difference of 3.124 mins
\end{lstlisting}

These are similar to the results of Poisson models. Moreover, we analyzed the data with the zero-inflated models. At first, we run the \textbf{zero-inflated negative binomial model} by the following: 

\begin{lstlisting}[language=R]
R> dfzinb(zinb100_Data)   
\end{lstlisting}

Here is the results: 
\begin{lstlisting}[language=R]
This is a (1, 3, 4, 2) model.

AIC value is 29.7.

Factors and variables in each factor:
  Factor1 Factor2 Factor3 Factor4
1 Var1    Var2    Var5    Var9   
2         Var4    Var7    Var10  
3         Var3    Var8           
4                 Var6           

Estimated value of r for the negative binomial distributed observations(s):
  Factor1 Factor2 Factor3 Factor4
1 4.7046  6.139   8.4336  6.5025 
2         0.4533  4.2253  10.1285
3         5.9512  0.5583         
4                 2.8136         

Estimated value of p for the negative binomial distributed observations(s):
  Factor1 Factor2 Factor3 Factor4
1 0.804   0.7556  0.8302  0.967  
2         0.4433  0.6859  0.9183 
3         0.7892  0.4753         
4                 0.6242         

Estimated parameters for the zero inflated part in the negative binomial 
distributed factor:
0 0.0233 0 0.197 0 0 0 0 0 0

Estimated value of r for the negative binomial distributed factor(s):
0 1.24 0.361 26.1 0 0 0 0 0 0

Estimated value of p for the negative binomial distributed factor(s):
0 0.656 0.466 0.922 0 0 0 0 0 0

Timing:
Time difference of 6.114 mins
\end{lstlisting}

Now, we consider the \textbf{zero-inflated truncated negative binomial model}: 

\begin{lstlisting}[language=R]
R> dfzinbt(zinb100_Data,6)   
\end{lstlisting}
Here are the results for this model with the truncation at 6. 
\begin{lstlisting}[language=R]
Call:
dfzinbt(y = zinb100_Data, A = 6)

This is a (1, 3, 4, 2) model.

AIC value is 29.5.

Factors and variables in each factor:
  Factor1 Factor2 Factor3 Factor4
1 Var1    Var2    Var5    Var9   
2         Var4    Var7    Var10  
3         Var3    Var8           
4                 Var6           

Estimated value of r for the negative binomial distributed observations(s):
  Factor1 Factor2 Factor3 Factor4
1 3.4935  1.5681  1.4556  45.3851
2         0.3644  4.6072  7.6677 
3         2.8035  0.4886         
4                 0.9588         

Estimated value of p for the negative binomial distributed observations(s):
  Factor1 Factor2 Factor3 Factor4
1 0.7568  0.4367  0.4861  0.9946 
2         0.3668  0.6845  0.8962 
3         0.6424  0.4247         
4                 0.3569         

Estimated parameters for the zero inflated part 
in the negative binomial distributed factor:
0.143

Estimated value of r for the negative binomial distributed factor(s):
1.04 0.326 4.13

Estimated value of p for the negative binomial distributed factor(s):
0.621 0.426 0.645

Timing:
Time difference of 15.87 mins
\end{lstlisting}

By comparing AIC of the eight different models, we can easily conclude that all models except the truncated Poisson, provided very similar results. Now we run a traditional factor analysis using the \texttt{factanal()} function in R. We obtained the following results:

\begin{lstlisting}[language=R]
R> Nfacs <- 3
R> factanal(zinb100_Data, Nfacs, rotation = "promax")

Call:
factanal(x = zinb100_Data, factors = Nfacs, rotation = "promax")

Uniquenesses:
 Var1  Var2  Var3  Var4  Var5  Var6  Var7  Var8  Var9 
0.936 0.805 0.785 0.518 0.797 0.855 0.271 0.639 0.024 
Var10 
0.258 

Loadings:
      Factor1 Factor2 Factor3
Var1   0.200           0.164 
Var2           0.111   0.449 
Var3           0.117   0.461 
Var4                   0.702 
Var5   0.141   0.392         
Var6           0.306  -0.174 
Var7           0.870   0.114 
Var8           0.578         
Var9   0.986                 
Var10  0.863                 

               Factor1 Factor2 Factor3
SS loadings      1.791   1.369   0.995
Proportion Var   0.179   0.137   0.100
Cumulative Var   0.179   0.316   0.416

Factor Correlations:
        Factor1 Factor2 Factor3
Factor1  1.0000 -0.0562  0.0441
Factor2 -0.0562  1.0000 -0.2015
Factor3  0.0441 -0.2015  1.0000

Test of the hypothesis that 3 factors are sufficient.
The chi square statistic is 9.14 on 18 degrees of freedom.
The p-value is 0.957     
\end{lstlisting}

It can be easily observed that traditional factor analysis failed to identify the true model. Moreover, the results are difficult to interpret. One justification is that the data set are not normally distributed. But our discrete factor analysis identified the correct model.  

\subsection{Example 2: Survey data}

In this subsection, we fit the models using a real dataset. The data set is available at \url{http://openpsychometrics.org/_rawdata/}. The questionnaire is five Likert scale data (from strongly disagree (1) to strongly agree (5)). For some negative questions, the scale was reversed. We choose 13 variables (Optimism Scale: opt1-opt6)  Mastery Scale (mast1-mast7) to identity the hidden factors. First we subtract $1$ from the data to fit the truncated Poisson model (Truncation is 4 here). The results are given below:

\begin{lstlisting}[language=R]
R> dfs <- survey_data - 1
R> dfpt(dfs, 4) 
Call:
dfpt(y = dfs, A = 4)

This is a (2, 3, 3, 3, 2) model.

AIC value is 33.1.

Factors and variables in each factor:
  Factor1 Factor2 Factor3 Factor4 Factor5
1 op1     op2     op3     mast1   mast6  
2 op5     op4     mast2   mast3   mast7  
3         op6     mast5   mast4          

Estimated parameters for each variable within each factor:
  Factor1 Factor2 Factor3 Factor4 Factor5
1 3.1502  0.6278  1.1509  0.479   0.6097 
2 2.9471  0.6454  1.2813  0.4432  1.0897 
3         1.269   2.4161  0.9195         

Estimated parameters for factors:
1.65 0.677 2.13 0.532 0.486
\end{lstlisting}

Now we run the traditional factor analysis by the following:

\begin{lstlisting}[language=R]
R>factanal(dfs, 5, rotation="promax")$loadings

Loadings:
      Factor1 Factor2 Factor3 Factor4 Factor5
op1    0.543   0.186  -0.240                 
op2                    0.680                 
op3    0.665          -0.142   0.142         
op4   -0.155           0.676                 
op5    0.406   0.172          -0.271  -0.157 
op6                            1.007         
mast1          0.438   0.249                 
mast2  0.547  -0.174   0.151                 
mast3          0.568                   0.114 
mast4          0.790                         
mast5  0.407  -0.231   0.148                 
mast6                                  0.773 
mast7                                  0.595 

               Factor1 Factor2 Factor3 Factor4 Factor5
SS loadings      1.412   1.308   1.129   1.116   1.007
Proportion Var   0.109   0.101   0.087   0.086   0.077
Cumulative Var   0.109   0.209   0.296   0.382   0.459
\end{lstlisting}

We observe that there are some similarities between the results of the traditional factor analysis and the discrete factor analysis using truncated Poisson model. However, the results of the truncated Poisson model are much easier to interpret.  To save space, we don't run the other discrete models here.

\subsection{Example 3: Sexual Compulsivity Scale (SCS data)}

\begin{figure}[h]
    \centering    
    \includegraphics[scale=.5]{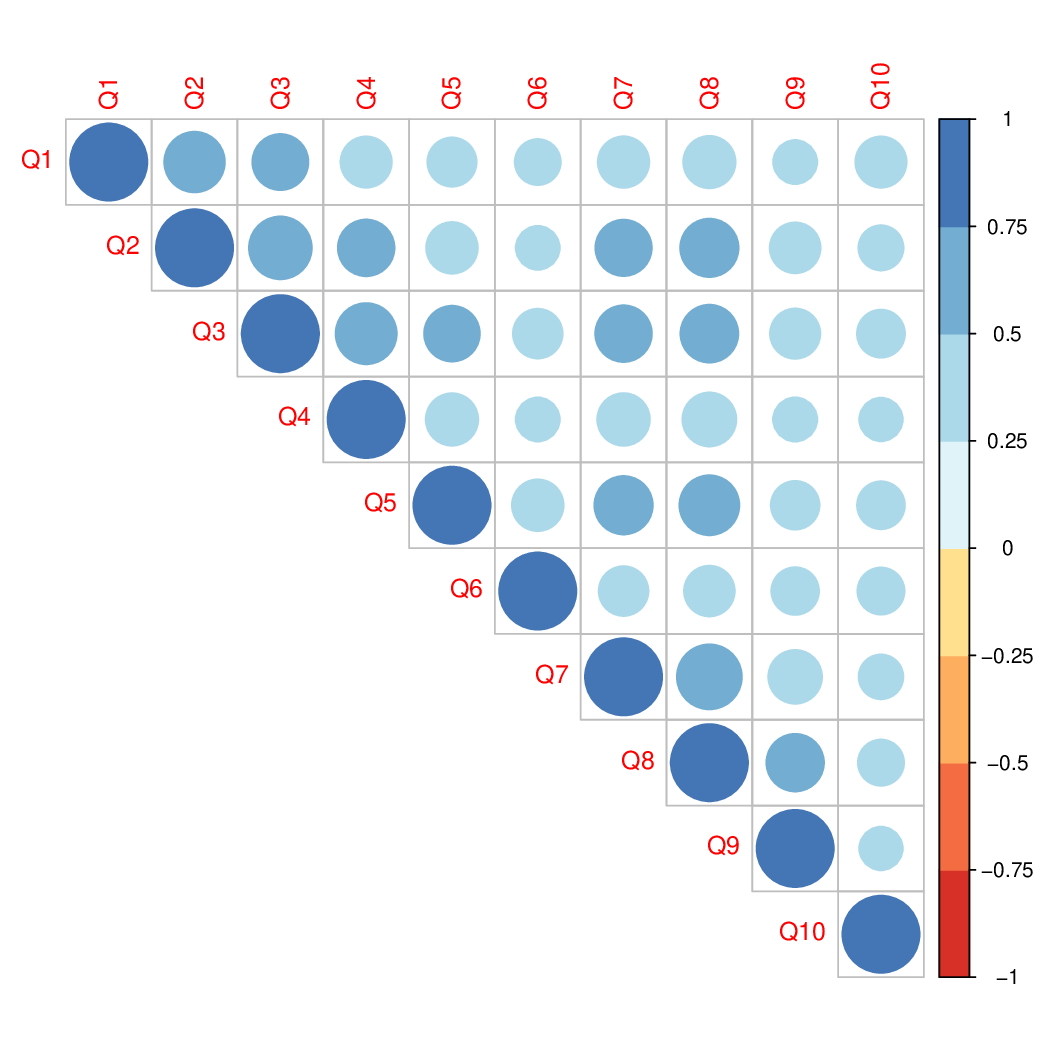}
    \caption{Correlation Matrix for the Sexual Compulsivity Scale (SCS data)}
    \label{scs_cor_data}
\end{figure}

In this subsection, we fit the models using Sexual Compulsivity Scale (SCS) dataset. The SCS was developed to assess tendencies toward sexual preoccupation
and hypersexuality. Items were initially derived from self-descriptions of persons who self-identify as
having a ‘sexual addiction’. The self-descriptors were taken from a brochure for a sexual addictions
self-help group. The scale should predict rates of sexual behaviors, numbers of sexual
partners, practice of a variety of sexual behaviors, and histories of sexually transmitted diseases. The
scale is internally consistent with Alpha coefficients that range between $0.85$ and $0.91$. The data set available at \url{http://openpsychometrics.org/_rawdata/} and updated at 7/16/2012. The data set has 3376 observations with 13 variables of which 10 variables were the questions about the description of the sexual behaviours. The full description of the data and motivation can be found in \citet{kalichman2001sexual}. The correlation matrix for this data set is given in Figure \ref{scs_cor_data}.

It is observed that there are some highly correlated variables suggesting some unknown factors associated with the self identified sexual behaviours. The results of discrete factor analysis using the Poisson model are given below:

\begin{lstlisting}[language=R]
R> dfp(SCS_data) 

Call:
dfp(y = SCS_data)

This is a (5, 5) model.

AIC value is 28.2.

Factors and variables in each factor:
  Factor1 Factor2
1 Q1      Q5     
2 Q2      Q7     
3 Q3      Q8     
4 Q4      Q6     
5 Q10     Q9     

Estimated parameters for each variable within each factor:
  Factor1 Factor2
1   0.913   0.697
2   0.832   0.650
3   0.827   0.748
4   0.541   1.546
5   1.116   0.915

Estimated parameters for factors:
1.39 1.54

Timing:
Time difference of 2.512 mins
\end{lstlisting}

Next, we run the traditional factor analysis as follows:

\begin{lstlisting}[language=R]
R> Nfacs = 2
R> factanal(SCS_data, Nfacs, rotation = "promax")$loadings

Loadings:
    Factor1 Factor2
Q1   0.727         
Q2   0.787         
Q3   0.809         
Q4   0.673         
Q5   0.137   0.594 
Q6   0.149   0.415 
Q7           0.795 
Q8           0.897 
Q9           0.604 
Q10  0.364   0.156 

               Factor1 Factor2
SS loadings      2.432   2.355
Proportion Var   0.243   0.235
Cumulative Var   0.243   0.479
\end{lstlisting}
It is seen that similar results to the discrete factor analysis emerged.  However, we run the truncated Poisson mode with \texttt{dfpt()} function as follows:

\begin{lstlisting}[language=R]
R> dfpt(SCS_data, 4)
\end{lstlisting}

\begin{lstlisting}[language=R]
Call:
dfpt(y = SCS_data, A = 4)

This is a (6, 4) model.

AIC value is 26.5.

Factors and variables in each factor:
  Factor1 Factor2
1 Q1      Q6     
2 Q2      Q7     
3 Q3      Q8     
4 Q4      Q9     
5 Q5             
6 Q10            

Estimated parameters for each variable within each factor:
  Factor1 Factor2
1 1.1696  0.7294 
2 1.1628  0.8564 
3 0.8036  1.0938 
4 1.1842  2.4547 
5 1.5892         
6 2.5092         

Estimated parameters for factors:
1.7 3

Timing:
Time difference of 9.959 mins
\end{lstlisting}
The comparison of the Poisson model to the truncated Poisson model reveals that the truncated Poisson model performs better than the usual Poisson model. 

\subsection{Example 4:  Agriculture, damage to potato tubers}

In this subsection, we repeat the analysis of Example 4 which is available in the \citet{larsson2021applications}. They only considered the four variables in the dataset (from \cite{wright2013agridat}) that are either ordinal or counts. These variables are $x_1$: Energy factor in the ordinal scale of $(1, 2)$, $x_2$: weight in the ordinal scale of $(1-3)$, $x_3$: damage category in the ordinal scale of $1-4$, and $x_4$: the count of tubers in each combination of categories as an integer value. In order to generate  positive correlations, we apply the same transformation in \citet{larsson2021applications} as follows:
\begin{eqnarray*}
 y_1&=&x_1-min(x_{i1})\\
 y_j&=&max(x_{ij})-x_j\\
 y_4&=&x_4
\end{eqnarray*}
for $i=1, 2, \ldots, n$ and $j=2,3$.
Figure \ref{potato_cor} depicts the correlations between the variables $y_1, y_2, y_3, y_4$. It is evident that the variables $y_3$ and $y_4$ are highly correlated. 

Now, we proceed with the discrete factor analysis using the Poisson model by the following:

\begin{figure}[ht]
    \centering
    \includegraphics[width=0.5\textwidth]{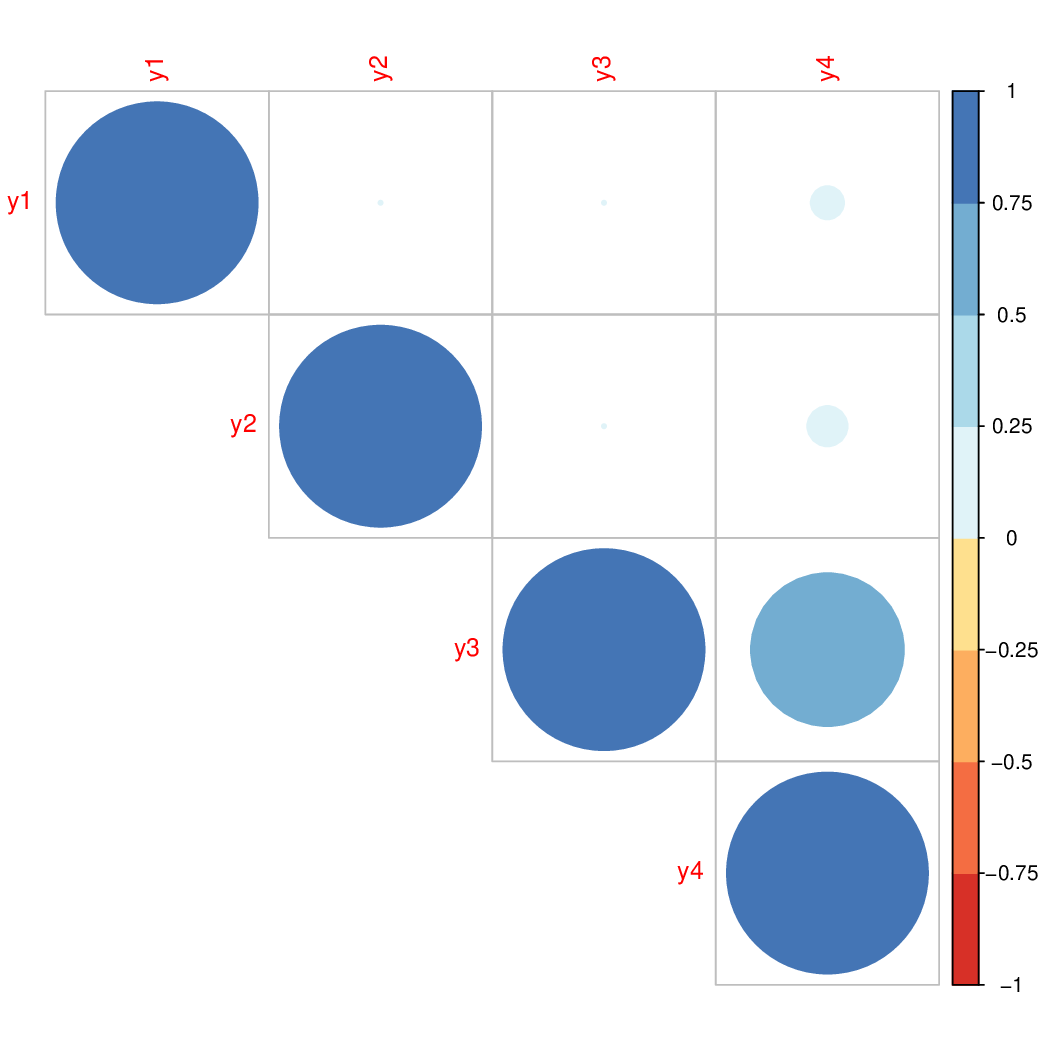}
    \caption{Correlation plot for the Agriculture, damage to potato tubers}
    \label{potato_cor}
\end{figure}
\begin{lstlisting}[language=R]
R> dfp(potato_data)
\end{lstlisting}
\begin{lstlisting}[language=R]
Call:
dfp(y = potato_data)

This is a (1, 1, 2) model.

AIC value is 16.

Factors and variables in each factor:
  Factor1 Factor2 Factor3
1 y1      y2      y3     
2                 y4     

Estimated parameters for each variable within each factor:
  Factor1 Factor2 Factor3
1 0.5     1       0.8687 
2                 4.0441 

Estimated parameters for factors:
0.631

Timing:
Time difference of 1.179 secs
\end{lstlisting}
As we expected from the correlation plot (Figure \ref{potato_cor}), the discrete  Poisson model identified one common factor in the data which is exactly the same as the one given in \citet{larsson2021applications}. 

We also run the zero-inflated Poisson model as below:

\begin{lstlisting}[language=R]
R> dfzip(potato_data)    
\end{lstlisting}

The results are obtained as follows: 

\begin{lstlisting}[language=R]
This is a (1, 1, 2) model.

AIC value is 12.7.

Factors and variables in each factor:
  Factor1 Factor2 Factor3
1 y1      y2      y3     
2                 y4     

Estimated zero-inflated parameters for each variable within each factor:
  Factor1 Factor2 Factor3
1 0       0       0      
2                 0.472  

Estimated parameters for each variable within each factor:
  Factor1 Factor2 Factor3
1 0.5     1       0.8463 
2                 7.616  

Estimated zero-inflated parameters for each factor:
0.445

Estimated parameters for factors:
1.18

Timing:
Time difference of 43.71 secs   
\end{lstlisting}
It is observed that the zero-inflated Poisson model provided lower AIC compared to the usual Poisson model. This is expected from the the Figure \ref{potato_cor} as $y_4$ seems to be a zero-inflated variable. 

\section{Limitation of the discFA Package}\label{sec:limit}

The \texttt{discFA} package can run the factor analysis for many discrete data sets. There is no specific issue except highly negatively correlated variables (less than -0.50). For instance, there was a high negative correlation  ($-0.58$) between two variables in the last example ($x_3,x_4$) which turned out to have an independent model given below (with the warning message):
\begin{lstlisting}[language=R]
R> dfp(original_potato_data)
There are some highly negative correlations among some variables
so the findings may not be stable.

Call:
dfp(y = original_potato_data)

Independent model!

This is a (1, 1, 1, 1) model.

AIC value is 17.69418.

Factors and variables in each factor:
  Factor1 Factor2 Factor3 Factor4
1 energy  weight  damage  count  

Estimated parameters (mu) for each variable within each factor:
  Factor1 Factor2 Factor3 Factor4
1     1.5       2     2.5  4.6753

Estimated Parameter (Lambda) for Factors:
0 0 0 0

Timing:
Time difference of 0.5549 secs    
\end{lstlisting}

There are some remedies for that as have been described in \citet{larsson2021applications} for example, reversing the scale which we performed in that example and obtained the correct model. Another limitation is the negative values in the data which is not so common in the Likert scale data. Finally, if there are some missing data, they could be addressed in advance by using the multiple imputation of \citet{white2011multiple} or some other machine learning imputation methods. Complete case analysis is another approach that can be applied if the portion of missingness is low ($<0.05$).    

\section{Summary and discussion} \label{sec:summary}
Traditional factor analysis often falls short when dealing with count data due to its underlying assumptions. Discrete factor analysis (DFA) emerges as a powerful and appropriate alternative, directly addressing the unique characteristics of non-negative, discrete observations mainly, for the Likert scale data. By explicitly modeling these data with distributions like the Poisson and negative binomial, and accounting for common complexities such as zero-inflation and truncation, DFA provides a more accurate and robust framework for uncovering latent structures. The \texttt{discFA} package in \texttt{R}, with its efficient forward search algorithm and reliance on AIC for model selection, simplifies the application of these sophisticated methods. As demonstrated through diverse examples across psychology, agriculture, and the automotive industry, DFA offers a statistically sound approach to analyzing count data, yielding more reliable insights than traditional methods could provide. Researchers in various fields can confidently adopt DFA to gain a deeper understanding from their count-based datasets. For future work, we aim to propose confirmatory factor analysis for discrete data in a new contribution. Further, the approaches we proposed here can be applied to skewed continuous data which is the subject of another contribution.


\bigskip
\noindent\textbf{Consent to participate}
Not applicable.
\newline
\noindent\textbf{Consent for publication}
Not applicable.
\newline
\noindent\textbf{Funding} Not Applicable.
\newline
\noindent\textbf{Data Availability} The dataset supports the findings of this study are openly available in reference list also in the discFA package.\\
\url{https://cran.r-project.org/web/packages/discFA/index.html}
\newline
\noindent\textbf{Declarations}
The authors declare that they have no financial interests.
\newline
\noindent\textbf{Conflict of interest} All authors declare that they have no conflict of interest.
\newline
\noindent\textbf{Ethics statements} 
Not applicable.
\newline
\bigskip

\bibliography{References}
\begin{appendix}
\section{Appendix: Proof Proposition 1}
Let the pmf of $X$ be given by
\begin{equation}
f(x;\theta)=h(x)\exp\{x ~ a(\theta)-b(\theta)\}.\label{pmfx}
\end{equation}
Denoting the derivative of $a(\theta)$ w.r.t. $\theta$ by $a'(\theta)$, and similarly for $b(\theta)$, it follows readily that
\begin{equation}
E(X)=\frac{b'(\theta)}{a'(\theta)}.\label{EX}
\end{equation}
Now, consider $Y_j=U+X_j$ for $j=1,...,m$, where the random variables $U,X_1,...,X_m$ are simultaneously independent, all having density $f(x;\theta_j)$ as in (\ref{pmfx}), where $j=0$ corresponds to $U$. 

Say that we have observations $(y_{11},...,y_{1m}),\ldots,(y_{n1},,...,y_{nm})$.
Letting $z_i=\min(y_{i1},...,y_{im})$, from (\ref{pmfx}), the likelihood is
\begin{align}
L(\theta_0,\theta_1,...,\theta_m)
&=\prod_{i=1}^n\left[\sum_{u_i=0}^{z_i} 
f(u_i;\theta_0)f(y_{i1}-u_i;\theta_1)\ldots f(y_{im}-u_i;\theta_m)
\right]\notag\\
&=\prod_{i=1}^n\left[\sum_{u_i=0}^{z_i} 
h(u_i)h(y_{i1}-u_i)\ldots h(y_{im}-u_i)
\cdot\exp\left\{-\sum_{j=0}^m b(\theta_j)\right\}\right.\notag\\
&\left.
\times\exp\left\{u_ia(\theta_0)+\sum_{j=1}^m (y_{ij}-u_i)a(\theta_j)\right\}
\right]\notag\\
&=\exp\left\{-n\sum_{j=0}^m b(\theta_j)\right\}
\sum_{u_1=0}^{z_1}\ldots\sum_{u_n=0}^{z_n}
\left\{\prod_{i=1}^n h(u_i)\right\}\exp\left\{\sum_{i=1}^nu_i a(\theta_0)\right\}\notag\\
&\times g_1(\theta_1) \ldots g_m(\theta_m),\label{likp1}
\end{align}
where for $j=1,...,m$,
\begin{align}
g_j(\theta_k)&=\left\{\prod_{i=1}^n h(y_{ij}-u_i)\right\}
\exp\left\{\sum_{i=1}^n(y_{ij}-u_i)a(\theta_j)\right\}\notag\\
&=C_j\exp\left\{\left(n\bar y_j-\sum_{i=1}^n u_i\right)a(\theta_j)\right\},
\label{gj}
\end{align}
with
$$C_j=\prod_{i=1}^nh(y_{ij}-u_i).$$
Now, suppressing the arguments of the likelihood function in (\ref{likp1}), its first derivative w.r.t. $\theta_1$ may be expressed as
\begin{align*}
\frac{\partial L}{\partial\theta_1}&=-nb'(\theta_1)L
+\exp\left\{-n\sum_{j=0}^m b(\theta_j)\right\}\sum_{u_1=0}^{z_1}\ldots\sum_{u_n=0}^{z_n}
\left\{\prod_{i=1}^n h(u_i)\right\}\\
&\times\exp\left\{\sum_{i=1}^nu_i a(\theta_0)\right\}
 g_1'(\theta_1)g_2(\theta_2)\ldots g_m(\theta_m),
\end{align*}
where via (\ref{gj}),
$$g_1'(\theta_1)=\left(n\bar y_1-\sum_{i=1}^n u_i\right)a'(\theta_1)
g_1(\theta_1),$$
so that
\begin{equation}
\frac{\partial L}{\partial\theta_1}=\left\{-b'(\theta_1)+\bar y_1a'(\theta_1)\right\}nL-a'(\theta_1)\exp\left\{-n\sum_{j=0}^m b(\theta_j)\right\}\sum_{i=1}^n A_i,\label{likder}
\end{equation}
where e.g.
\begin{align}
A_n&=\sum_{u_1=0}^{z_1}\ldots\sum_{u_{n-1}=0}^{z_{n-1}}
\left\{\prod_{i=1}^{n-1} h(u_i)\right\}
\exp\left\{\sum_{i=1}^{n-1}u_i a(\theta_0)\right\}\notag\\
&\times\sum_{u_n=0}^{z_n}h(u_n)u_n\exp\left\{u_na(\theta_0)\right\}
g_1(\theta_1)\ldots g_m(\theta_m).\label{An}
\end{align}
Now, because
$$\frac{\partial}{\partial\theta_0}\exp\left\{u_na(\theta_0)\right\}
=a'(\theta_0)\exp\left\{u_na(\theta_0)\right\},$$
(\ref{An}) and (\ref{likp1}) yield
\begin{align*}
\sum_{i=1}^n A_i
&=\frac{1}{a'(\theta_0)}\frac{\partial}{\partial\theta_0}
\sum_{u_1=0}^{z_1}\ldots\sum_{u_n=0}^{z_n}\left\{\prod_{i=1}^n h(u_i)\right\}\exp\left\{\sum_{i=1}^n u_ia(\theta_0)\right\}
g_1(\theta_1)\ldots g_m(\theta_m)\\
&=\frac{1}{a'(\theta_0)}\frac{\partial}{\partial\theta_0}
\left[\exp\left\{n\sum_{j=0}^m b(\theta_j)\right\}L\right]\\
&=\frac{1}{a'(\theta_0)}\exp\left\{n\sum_{j=0}^m b(\theta_j)\right\}
\left\{b'(\theta_0)nL+\frac{\partial L}{\partial\theta_0}\right\},
\end{align*}
so that by (\ref{likder}),
\begin{equation}
\frac{\partial L}{\partial\theta_1}=\left\{-b'(\theta_1)+\bar y_1a'(\theta_1)\right\}nL-\frac{a'(\theta_1)}{a'(\theta_0)}
\left\{b'(\theta_0)nL+\frac{\partial L}{\partial\theta_0}\right\}.\label{likder1}
\end{equation}
Hence, because $\partial L/(\partial\theta_0)=0$ for the MLE $\theta_0=\hat\theta_0$, (\ref{likder1}) yields
$$\left.\frac{\partial L}{\partial\theta_1}\right|_{\theta_0=\hat\theta_0}
=\left\{-\frac{b'(\theta_1)}{a'(\theta_1)}+\bar y_1-\frac{b'(\theta_0)}{a'(\theta_0)}\right\}a'(\theta_1)nL,$$
i.e. from (\ref{EX}) and $Y_1=U+X_1$,
$$\left.\frac{\partial L}{\partial\theta_1}\right|_{\theta_0=\hat\theta_0}
=\left\{-E(X_1)+\bar y_1-E(U)\right\}a'(\theta_1)nL=\left\{\bar y_1-E(Y_1)\right\}a'(\theta_1)nL,$$
which, since $a(\theta_1)$ is not constant, is zero only if $E(Y_1)=\bar y_1$, as was to be shown. The proof that at the MLEs, $E(Y_j)=\bar y_j$ for any $j>1$, is similar.



\end{appendix}

\end{document}